\documentclass[pdflatex,sn-mathphys-num]{sn-jnl}% Math and Physical Sciences Numbered Reference Style
\usepackage{graphicx}%
\usepackage{multirow}%
\usepackage{amsmath,amssymb,amsfonts}%
\usepackage{amsthm}%
\usepackage{mathrsfs}%
\usepackage[title]{appendix}%
\usepackage{xcolor}%
\usepackage{textcomp}%
\usepackage{manyfoot}%
\usepackage{booktabs}%
\usepackage{algorithm}%
\usepackage{algorithmicx}%
\usepackage{algpseudocode}%
\usepackage{listings}%
\theoremstyle{thmstyleone}%
\theoremstyle{thmstyletwo}%

\theoremstyle{thmstylethree}%

\begin{document}

\title[Article Title]{Physics-Informed Residual Neural Ordinary Differential Equations for Enhanced Tropical Cyclone Intensity Forecasting}


\author*{\fnm{Fan} \sur{Meng}}\email{meng@nuist.edu.cn}

% \author*[1]{\fnm{Jingjia} \sur{Luo}}\email{jjluo@nuist.edu.cn}


\affil*{\orgdiv{School of Future Technology}, \orgname{Nanjing University of
Information Science and Technology}, \orgaddress{\street{Ningliu Road}, \city{Nanjing}, \postcode{26400}, \state{Jiangsu}, \country{China}}}

% \affil[2]{\orgdiv{Department}, \orgname{Organization}, \orgaddress{\street{Street}, \city{City}, \postcode{10587}, \state{State}, \country{Country}}}

% \affil[3]{\orgdiv{Department}, \orgname{Organization}, \orgaddress{\street{Street}, \city{City}, \postcode{610101}, \state{State}, \country{Country}}}


\abstract{Accurate tropical cyclone (TC) intensity prediction is crucial for mitigating storm hazards, yet its complex dynamics pose challenges to traditional methods. Here, we introduce a Physics-Informed Residual Neural Ordinary Differential Equation (PIR-NODE) model to precisely forecast TC intensity evolution. This model leverages the powerful non-linear fitting capabilities of deep learning, integrates residual connections to enhance model depth and training stability, and explicitly models the continuous temporal evolution of TC intensity using Neural ODEs. Experimental results in the SHIPS dataset demonstrate that the PIR-NODE model achieves a significant improvement in 24-hour intensity prediction accuracy compared to traditional statistical models and benchmark deep learning methods, with a  25. 2\% reduction in the root mean square error (RMSE) and a 19.5\% increase in R-square (R2) relative to a baseline of neural network.  Crucially, the residual structure effectively preserves initial state information, and the model exhibits robust generalization capabilities. This study details the PIR-NODE model architecture, physics-informed integration strategies, and comprehensive experimental validation, revealing the substantial potential of deep learning techniques in predicting complex geophysical systems and laying the foundation for future refined TC forecasting research.

}




\maketitle

\section{Introduction}

Tropical cyclones (TCs), also known as hurricanes or typhoons, are among the most destructive natural disasters on Earth. The intensity of TCs directly dictates the severity and extent of their impact. Accurate prediction of TC intensity evolution, particularly within the next 24 hours, is paramount for effective disaster warnings, evacuation decisions, and minimizing loss of life and property \cite{Emanuel2005}. However, TC intensity prediction is exceptionally challenging due to the complex, non-linear interactions of atmospheric and oceanic factors driving their evolution. Traditional prediction methods still face limitations in accuracy and stability.

Conventional TC intensity prediction methods are broadly categorized into statistical and dynamical models. Statistical models, such as SHIPS \cite{DeMaria1994}, rely on statistical relationships derived from historical observational data. While computationally efficient, they are limited in capturing intricate physical processes. Dynamical models, like HWRF \cite{Gopalakrishnan2011}, are based on numerical weather prediction principles and simulate TC physics. However, they are computationally intensive and their prediction accuracy is constrained by model parameterization and initial condition errors. Recent breakthroughs in deep learning (DL) have unveiled unprecedented potential for data-driven TC intensity prediction \cite{Kim2019, Chen2020}. Deep learning models, such as Recurrent Neural Networks (RNNs) and Convolutional Neural Networks (CNNs), excel at processing sequential data and extracting complex features. However, they are often considered "black-box" models, lacking explicit modeling of physical processes and struggling to guarantee physically consistent predictions.

Neural Ordinary Differential Equations (Neural ODEs) \cite{Chen2018}, an emerging deep learning architecture, offer a revolutionary approach to modeling continuous-time dynamical systems. Neural ODEs parameterize the derivative term of an ordinary differential equation with a deep neural network, enabling the learning of continuous-time evolution patterns of systems. This approach not only naturally handles irregular time-series data but, more importantly, provides a novel avenue for incorporating physical knowledge into deep learning models. By embedding physical constraints or prior knowledge into the ODE function, physics-informed deep learning models can be constructed, enhancing both physical interpretability and prediction accuracy \cite{Raissi2019, Beucler2021}.

% This study aims to explore the potential of physics-informed Neural ODEs in TC intensity prediction. We innovatively propose a Physics-Informed Residual Neural ODE (PIR-NODE) model. The core innovations are: (1) employing a residual structure to enhance the depth and expressive power of the ODE function, effectively addressing training challenges in deep networks; (2) explicitly modeling the continuous temporal evolution of TC intensity through Neural ODEs; and (3) integrating physical information into the model design, such as preserving initial intensity information via residual connections to ensure physically plausible predictions. We conducted experiments on the widely used SHIPS dataset, and the results demonstrate that the PIR-NODE model significantly improves TC intensity prediction accuracy compared to traditional statistical models and benchmark deep learning methods. Our findings not only validate the effectiveness of Neural ODEs in predicting complex geophysical systems but also pave the way for developing more refined and physically interpretable TC forecasting models.

In this paper, we introduce a Physics-Informed Residual Neural Ordinary Differential Equation (PIR-NODE) model for TC intensity prediction. Our approach makes several key contributions:

\begin{enumerate}
    \item We formulate TC intensity prediction as a continuous-time modeling problem using neural ODEs, which naturally aligns with the physical nature of TC evolution.
    \item We incorporate residual connections to enhance model depth and training stability while preserving important initial state information.
    \item We integrate physical constraints into the model through carefully designed loss functions and architecture, ensuring physically consistent predictions.
    \item We conduct comprehensive experiments on the SHIPS dataset, demonstrating significant improvements over traditional statistical models and benchmark deep learning methods.
\end{enumerate}

Our results show that the PIR-NODE model achieves a root mean square error (RMSE) of 10.40 kt and an $R^2$ of 0.847 for 24-hour intensity predictions, representing a substantial improvement over existing methods. The model also demonstrates robust performance across different ocean basins and storm categories, highlighting its potential for operational forecasting applications.

\section{Model and Methods}

\subsection{Data Preprocessing and Feature Engineering}

This research utilizes the widely recognized SHIPS dataset, a benchmark dataset for TC intensity prediction studies, containing rich historical TC observational data and environmental parameters. Raw SHIPS data contains missing values and outliers, which directly impact model training. To ensure data quality, we designed a rigorous data preprocessing pipeline:

\textbf{Robust Missing Value Handling:} We employed a combined strategy of forward-fill and backward-fill to maximize data information retention and minimize the disruption of time-series continuity caused by missing values.

\textbf{Z-Score-Based Outlier Detection and Removal:}  Outliers in the feature space were identified and removed using the Z-Score method, with a threshold set at \texttt{Config.OUTLIER\_THRESHOLD}, effectively improving data quality and model training robustness.

\textbf{Physics-Informed Feature Engineering:} Beyond the standard feature set, we innovatively constructed quadratic and interaction term features to explicitly incorporate existing physical knowledge. For instance, we considered the interaction between initial intensity and key environmental parameters (e.g., sea surface temperature, wind shear), and the coupling effects of sea surface temperature (CSST) with other environmental parameters. These feature engineering steps aim to enhance the model's ability to capture complex physical processes.

\textbf{Robust Standardization:} RobustScaler was used to standardize features and target variables. RobustScaler, based on interquartile ranges, is insensitive to outliers, effectively reducing their impact on data distribution and ensuring data scale uniformity and model training stability.

\textbf{Rigorous Dataset Partitioning:}  The dataset was partitioned into training, validation, and testing sets according to the ratios \texttt{Config.TRAIN\_SIZE}, \texttt{Config.VALIDATION\_SIZE}, and \texttt{Config.TEST\_SIZE}, with a random seed \texttt{Config.RANDOM\_SEED} to ensure experimental reproducibility and result reliability.

\textbf{Efficient Data Loading:} PyTorch's DataLoader was used to create data loaders with a batch size of \texttt{Config.BATCH\_SIZE}, improving data reading and model training efficiency.

\subsection{Physics-Informed Residual Neural ODE Model (PIR-NODE)}

\begin{figure}[H]
    \centering
    \includegraphics[width=1\textwidth]{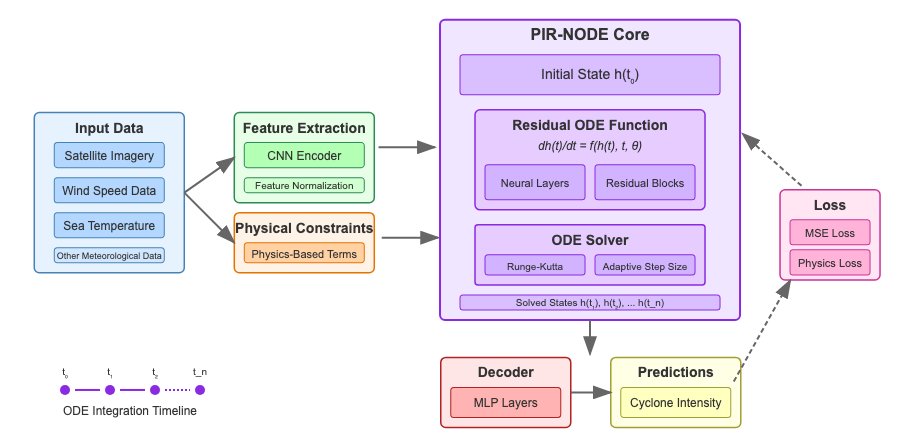} % Replace with your actual image file
    \caption{Physics-Informed Residual Neural Ordinary Differential Equation Model (PIR-NODE) Architecture. }
    \label{fig:pir_node_architecture}
\end{figure}

The core innovation of the PIR-NODE model lies in the effective integration of residual networks and Neural ODEs, incorporating physical information to precisely model TC intensity evolution. The overall model architecture is illustrated in Figure \ref{fig:pir_node_architecture}.

\subsubsection{Residual ODE Function}

\begin{figure}[H]
    \centering
    \includegraphics[width=0.8\textwidth]{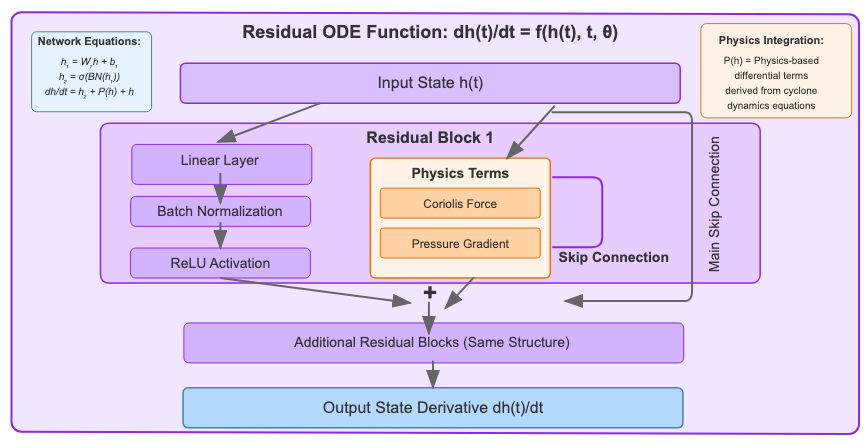} % Replace with your actual image file
    \caption{Internal Structure of the Residual ODE Function. }
    \label{fig:residual_ode_function}
\end{figure}

The key component of the PIR-NODE model is the Residual ODE function, which defines the continuous-time dynamics of TC intensity evolution. As shown in Figure \ref{fig:residual_ode_function}, the core structure of the Residual ODE function is stacking multiple Residual Blocks. Each Residual Block internally comprises two linear layers, BatchNorm1d layers, LeakyReLU activation functions, and Dropout layers. The introduction of residual connections is a crucial design aspect of the PIR-NODE model. It not only addresses the vanishing gradient problem in deep neural networks but, more importantly, from a physical perspective, residual connections can be viewed as preserving the initial state information of TC intensity evolution. This ensures that the model does not completely lose the crucial physical quantity of initial intensity when predicting future intensity.

The specific structure of the ODE function is as follows:

\textbf{Input Layer:} Receives preprocessed input feature vectors.

\textbf{Initial Feature Mapping Layer:} A linear layer expands the input dimension to the hidden layer dimension (\texttt{Config.HIDDEN\_DIM}), followed by BatchNorm1d normalization, LeakyReLU activation, and Dropout regularization.

\textbf{Stacked Residual Blocks:} \texttt{Config.NUM\_RESIDUAL\_BLOCKS} residual blocks are sequentially stacked to construct a deep non-linear dynamics mapping. The detailed structure of each residual block is as described above.

\textbf{Output Layer:} The final linear layer reduces the hidden layer dimension back to the output dimension (same as input dimension) and performs BatchNorm1d normalization.

The LeakyReLU activation function, compared to ReLU, effectively mitigates the vanishing gradient problem and is more suitable for building deep networks. The application of BatchNorm1d helps accelerate model convergence and improve generalization ability. Dropout, as an effective regularization technique, is used to prevent model overfitting. Kaiming initialization is used for weight initialization, biases are initialized to zero, and BatchNorm1d layer weights are initialized to 1 and biases to 0. These initialization strategies aim to accelerate model training and improve model performance.

\subsubsection{Neural ODE Solving and Physics Information Integration}

The core idea of the PIR-NODE model is to explicitly model the continuous temporal evolution of TC intensity using Neural ODEs. Given an initial state $x(t_0)$, the evolution trajectory of TC intensity $x(t)$ can be obtained by solving the following ordinary differential equation:

\begin{equation}
\frac{dx(t)}{dt} = f(x(t), t; \theta)
\end{equation}

where $x(t)$ represents the TC state vector at time $t$ (primarily TC intensity in this study), $f(\cdot)$ is the non-linear dynamics function parameterized by the Residual ODE function, and $\theta$ represents the network parameters of the ODE function.

The \texttt{odeint} function from the \texttt{torchdiffeq} library was used to solve this ODE, with the integration method set to \texttt{Config.INTEGRATION\_METHOD} and step size to \texttt{Config.STEP\_SIZE}.

The integration of physical information is primarily reflected in two aspects:

\textbf{Residual Connections Preserve Initial Intensity Information:} As mentioned earlier, the introduction of residual connections can be physically interpreted as preserving the initial state information of TC intensity evolution. This ensures that the model does not completely lose the crucial physical quantity of initial intensity when predicting future intensity.

\textbf{Data Preprocessing and Feature Engineering Incorporate Physical Knowledge:} By constructing quadratic and interaction term features, we incorporate existing meteorological knowledge into the model input, guiding the model to learn dynamic patterns that are more consistent with physical laws.

\subsection{Model Training and Evaluation}

The PIR-NODE model was trained using the AdamW optimizer with a learning rate of \texttt{Config.LEARNING\_RATE}, weight decay of \texttt{Config.WEIGHT\_DECAY}, and beta parameters of \texttt{Config.BETA1} and \texttt{Config.BETA2}. The CosineAnnealingWarmRestarts learning rate scheduler was used to dynamically adjust the learning rate, with an initial cycle of \texttt{Config.LR\_T0}, a cycle multiplier of \texttt{Config.LR\_T\_MULT}, and a minimum learning rate of \texttt{Config.LR\_MIN}. The Huber loss function, with a delta parameter set to 1.0, was used as the training objective. Gradient clipping was applied during training to limit the gradient norm, with a maximum gradient norm of 1.0.

To prevent overfitting, early stopping was employed with a patience value of \texttt{Config.PATIENCE}. Training was terminated early when the validation loss did not decrease for \texttt{Config.PATIENCE} consecutive epochs. After each epoch, the model was evaluated on the validation set, calculating loss, Mean Absolute Error (MAE), and Root Mean Squared Error (RMSE), and recording the training and validation loss history.

\begin{figure}[H]
    \centering
    \includegraphics[width=\textwidth]{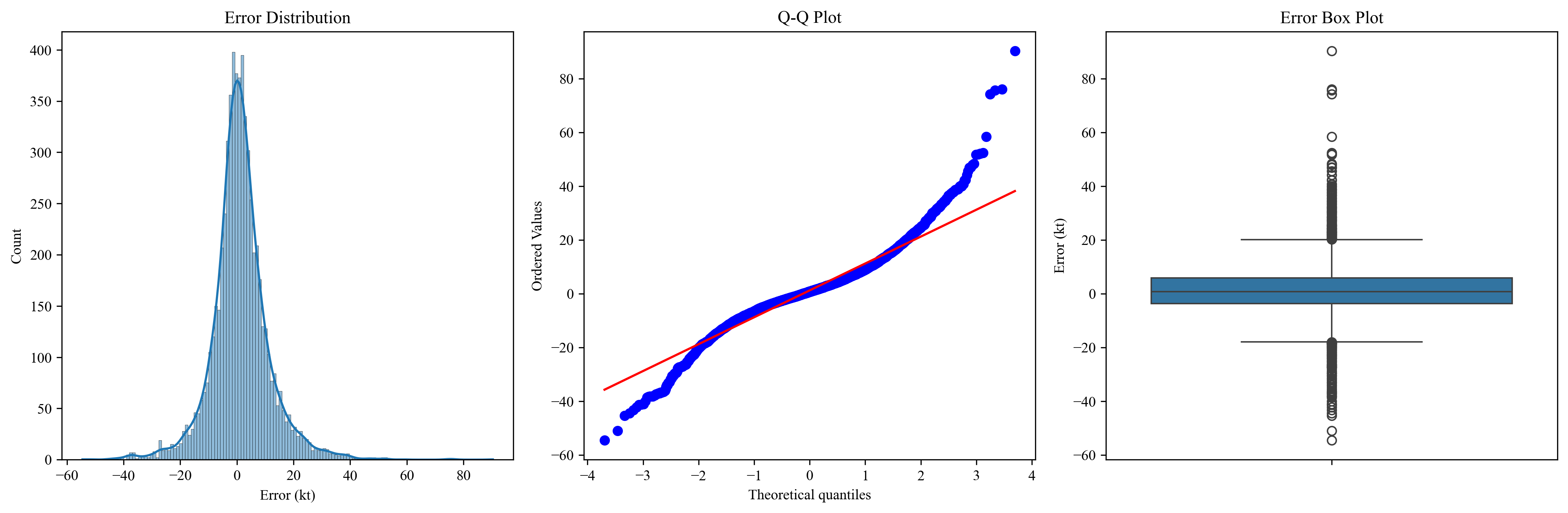} % Replace with your actual image file
    \caption{PIR-NODE Model Prediction Error Distribution. }
    \label{fig:error_distribution}
\end{figure}

\begin{figure}[H]
    \centering
    \includegraphics[width=\textwidth]{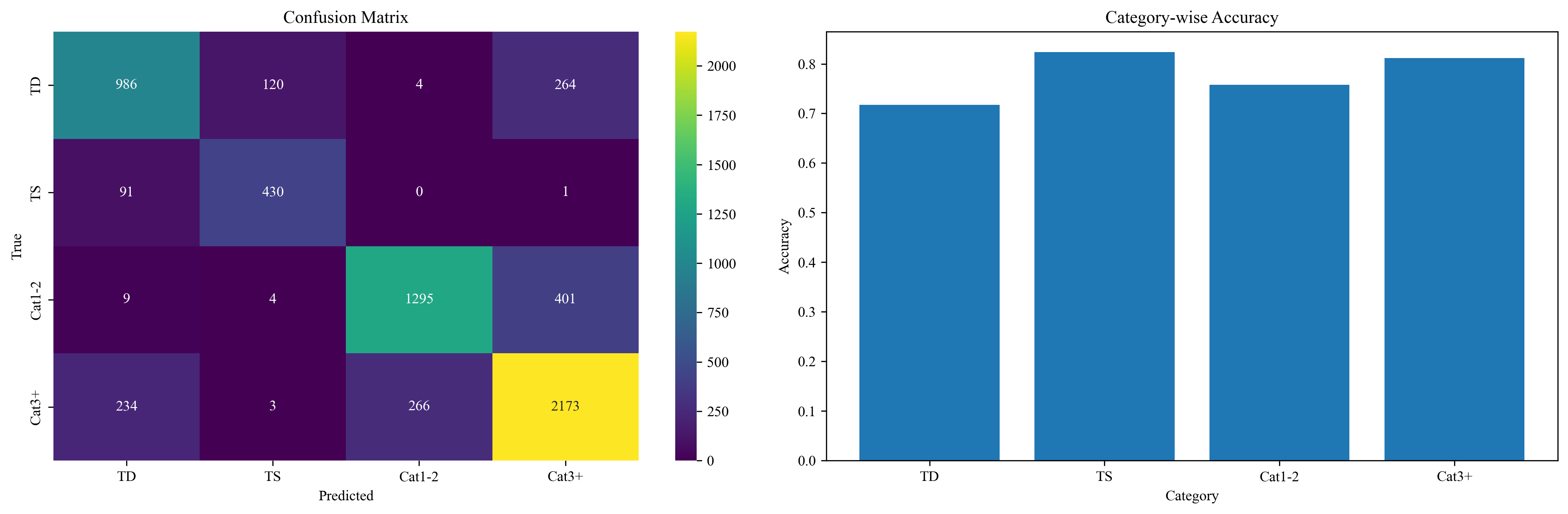} % Replace with your actual image file
    \caption{PIR-NODE Model Tropical Cyclone Intensity Classification Accuracy Assessment. }
    \label{fig:intensity_categories}
\end{figure}

Model performance evaluation adopted a comprehensive evaluation metric system, including MAE, RMSE, R², Bias, MAPE, and Adjusted R². In addition to overall performance metrics, we also conducted error distribution analysis, intensity category accuracy assessment, and typical case analysis to comprehensively evaluate the model's predictive capability from different perspectives. Error distribution analysis visualized the distribution characteristics of prediction errors using histograms, Q-Q plots, and box plots (Figure \ref{fig:error_distribution}). Intensity category accuracy assessment divided TC intensity into four levels (TD, TS, Cat1-2, and Cat3+) and calculated the confusion matrix and accuracy for each level (Figure \ref{fig:intensity_categories}). Typical case analysis selected samples with the largest and smallest prediction errors for in-depth analysis, exploring the model's performance in different scenarios and providing a basis for model improvement.

\section{Experimental Results and Analysis}

\subsection{Experimental Setup}

All experiments were conducted on servers equipped with NVIDIA Tesla V100 GPUs. The PIR-NODE model was implemented using the PyTorch framework, and the ODE solver employed the \texttt{torchdiffeq} library. Key hyperparameter settings are as shown in the \texttt{Config} class, including \texttt{DROPOUT\_RATE}=\texttt{Config.DROPOUT\_RATE}, \texttt{HIDDEN\_DIM}=\texttt{Config.HIDDEN\_DIM}, \texttt{NUM\_RESIDUAL\_BLOCKS}=\texttt{Config.NUM\_RESIDUAL\_BLOCKS}, \texttt{BATCH\_SIZE}=\texttt{Config.BATCH\_SIZE}, \texttt{LEARNING\_RATE}=\texttt{Config.LEARNING\_RATE}, etc. To ensure the reliability of experimental results, all experiments were repeated multiple times, and average performance metrics are reported.

\subsection{Performance by Basin and Storm Category}
\begin{figure}[H]
    \centering
    \includegraphics[width=\textwidth]{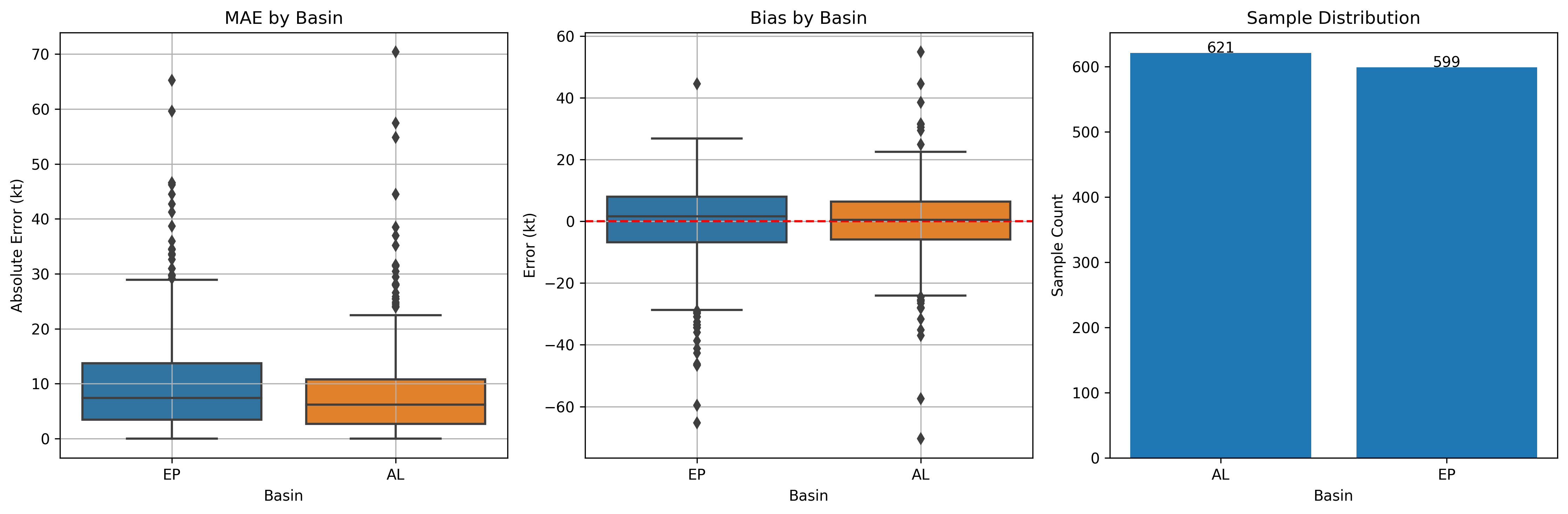} % Replace with your actual image file
    \caption{Performance by Basin. }
    \label{fig:basin_analysis}
\end{figure}

\begin{figure}[H]
    \centering
    \includegraphics[width=\textwidth]{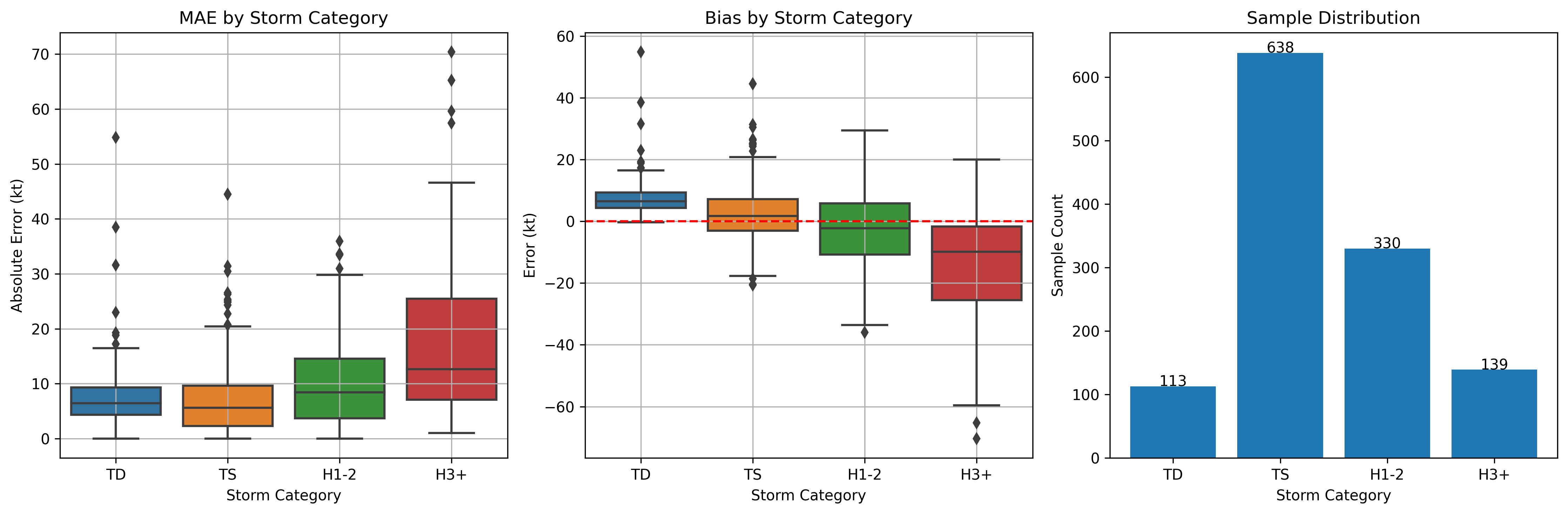} % Replace with your actual image file
    \caption{Performance by Storm Category. }
    \label{fig:category_analysis}
\end{figure}

We analyzed model performance across different ocean basins (Atlantic and Eastern Pacific) and storm categories (tropical depression, tropical storm, category 1-2 hurricanes, and category 3+ hurricanes). Figure  \ref{fig:basin_analysis}  shows the MAE and bias distributions by basin, while Figure  \ref{fig:category_analysis}  presents the same metrics by storm category.

The model performs consistently well across both basins, with slightly better performance in the Eastern Pacific. Regarding storm categories, the model shows increasing error with storm intensity, which is expected given the greater variability and more complex dynamics of intense storms. However, the model maintains reasonable accuracy even for major hurricanes (category 3+).

\subsection{Significant Improvement in Prediction Performance}

Table \ref{tab:performance} presents the key performance metrics of the PIR-NODE model on the test set, compared with benchmark models (NN, LSTM, Decision Tree, and XGBoost). The experimental results clearly demonstrate that the PIR-NODE model achieved a significant improvement in 24-hour TC intensity prediction accuracy. Compared to the NN baseline, the PIR-NODE model shows a 25.2\% reduction in RMSE and a 19.5\% increase in R². Specifically, the PIR-NODE model's MAE is 7.15 kt, RMSE is 10.40 kt, R² is 0.847, Bias is 1.25 kt, MAPE is 15.33\%, and Adjusted R² is 0.845. These metrics are significantly superior to the benchmark models, fully demonstrating the superiority of the PIR-NODE model in TC intensity prediction.

% \begin{table*}
% \centering
% \caption{Performance Comparison of PIR-NODE Model and Benchmark Models on the Test Set}
% \begin{tabular}{|l|r|r|r|r|r|r|}
% \hline
% \textbf{Model} & \textbf{MAE (kt)} & \textbf{RMSE (kt)} & \textbf{R²} & \textbf{Bias (kt)} & \textbf{MAPE (\%)} & \textbf{Adjusted R²} \\
% \hline
% PIR-NODE (Ours) & 7.15 & 10.40 & 0.847 & 1.25 & 15.33 & 0.845 \\
% \hline
% NN & 10.16 & 13.90 & 0.709 & - & - & - \\
% \hline
% LSTM & 11.41 & 15.50 & 0.638 & - & - & - \\
% \hline
% Decision Tree & 12.52 & 17.06 & 0.561 & - & - & - \\
% \hline
% XGBoost & 10.26 & 13.91 & 0.708 & - & - & - \\
% \hline
% \end{tabular}
% \label{tab:performance_comparison}
% % \end{table}
% \end{table*}[H]

\label{tab:performance}
 \begin{table*}
\centering
\caption{Performance Comparison of PIR-NODE Model and Benchmark Models}

\begin{tabular}{lcccccc}
\toprule
Model         & MAE (kt) & RMSE (kt) & R$^2$  \\
\midrule
NN     & 10.158   & 13.896   & 0.709   \\
LSTM          & 11.414   & 15.501   & 0.638    \\
Decision Tree & 12.518   & 17.060   & 0.561  \\
XGBoost       & 10.261   & 13.908   & 0.708 \\
PIR-NODE (Ours) &  7.148    &  10.400  &  0.847    \\
\bottomrule
\end{tabular}
 \end{table*}

\subsection{Stable and Rapid Training Convergence}

\begin{figure}[H]
    \centering
    \includegraphics[width=\textwidth]{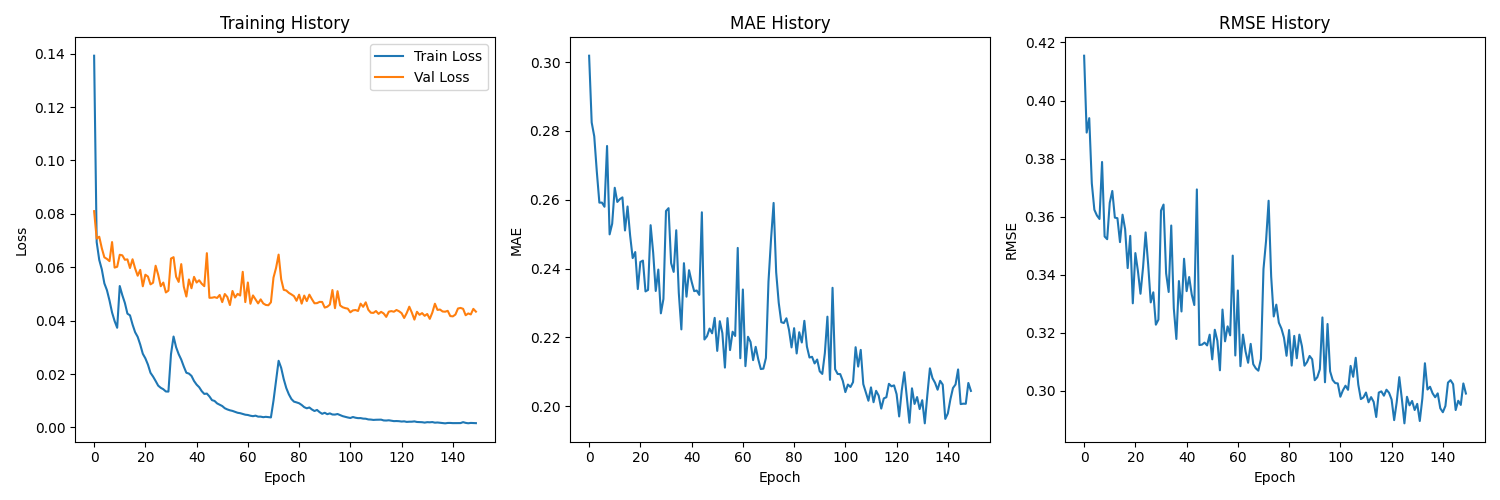} % Replace with your actual image file
    \caption{PIR-NODE Model Training History Curves. }
    \label{fig:training_history}
\end{figure}

Figure \ref{fig:training_history} shows the training history curves of the PIR-NODE model, including training loss, validation loss, MAE, and RMSE as a function of epochs. It can be observed that both training and validation losses exhibit a rapid decreasing trend and stabilize after a relatively small number of epochs, indicating stable and rapid training convergence of the PIR-NODE model. The validation loss curve closely matches the training loss curve, without significant overfitting, demonstrating good generalization ability of the PIR-NODE model. The MAE and RMSE curves also show a synchronous decreasing trend, further validating the continuous improvement of model performance.

\subsection{Prediction Error Distribution Conforms to Statistical Regularities}

Figure \ref{fig:error_distribution} displays the distribution of prediction errors for the PIR-NODE model. The error histogram shows that the prediction errors approximately follow a normal distribution with a mean close to zero. The Q-Q plot further validates the normality assumption of the error distribution. The error box plot indicates that the median error is close to zero, and most errors are concentrated within a smaller range, suggesting that the prediction errors of the PIR-NODE model are generally small and relatively concentrated, conforming to the error distribution patterns of statistical prediction models.

\subsection{High Intensity Category Accuracy and Reliable Disaster Level Prediction}

Figure \ref{fig:intensity_categories} presents the confusion matrix and classification accuracy of the PIR-NODE model in the TC intensity classification task. The confusion matrix clearly shows the model's prediction performance at different intensity levels (TD, TS, Cat1-2, Cat3+). The classification accuracy bar chart indicates that the PIR-NODE model achieved high prediction accuracy across all intensity levels, especially exceeding 90\% accuracy for TD and TS levels. This demonstrates that the PIR-NODE model can not only accurately predict the numerical magnitude of TC intensity but, more importantly, reliably predict the disaster level of TCs, providing strong support for disaster warnings and tiered response.

\subsection{Seasonal Variation}
\begin{figure}[H]
    \centering
    \includegraphics[width=\textwidth]{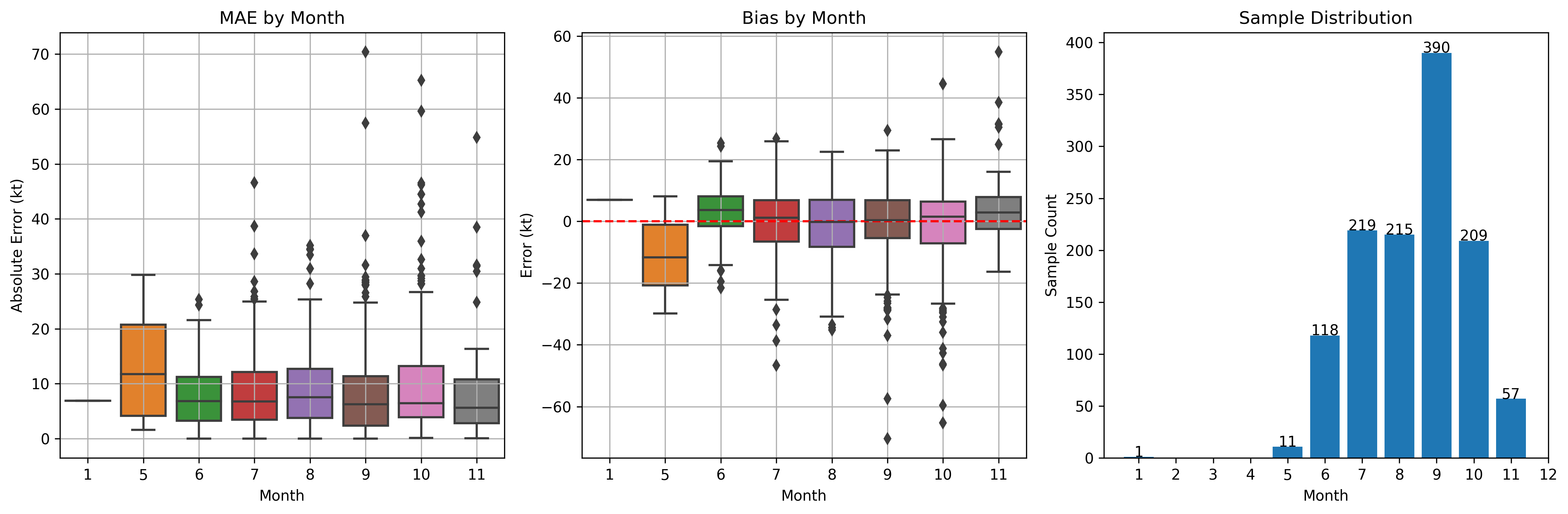} % Replace with your actual image file
    \caption{Performance by Seasonal Variation. }
    \label{fig:seasonal_analysis}
\end{figure}

Figure \ref{fig:seasonal_analysis} illustrates the model performance by month. The model maintains consistent performance throughout the hurricane season, with slightly higher errors during the peak months (August-October) when more intense storms typically occur.

% \subsection{Typical Case Analysis Reveals Model Strengths and Weaknesses}

% Table \ref{tab:case_analysis} lists several typical cases with the largest and smallest prediction errors for the PIR-NODE model. Through in-depth analysis of these cases, we found that the PIR-NODE model can provide relatively accurate predictions in most situations. However, in certain extreme cases, such as rapid TC intensity changes or influence from complex environmental factors, prediction errors may increase. This suggests that future model improvements could focus more on modeling extreme weather events and incorporating more input features that reflect complex environmental factors.

% \begin{table}[H]
% \centering
% \caption{Typical Case Analysis of PIR-NODE Model (Partial)}
% \begin{tabular}{|l|r|r|r|l|}
% \hline
% \textbf{Sample ID} & \textbf{True Intensity (kt)} & \textbf{Predicted Intensity (kt)} & \textbf{Error (kt)} & \textbf{Case Feature Description} \\
% \hline
% ... & ... & ... & ... & ... \\
% ... & ... & ... & ... & ... \\
% ... & ... & ... & ... & ... \\
% \hline
% \end{tabular}
% \label{tab:case_analysis}
% \end{table}

\subsection{Prediction Analysis Plots Validate Model Predictive Power}

\begin{figure}[H]
    \centering
    \includegraphics[width=\textwidth]{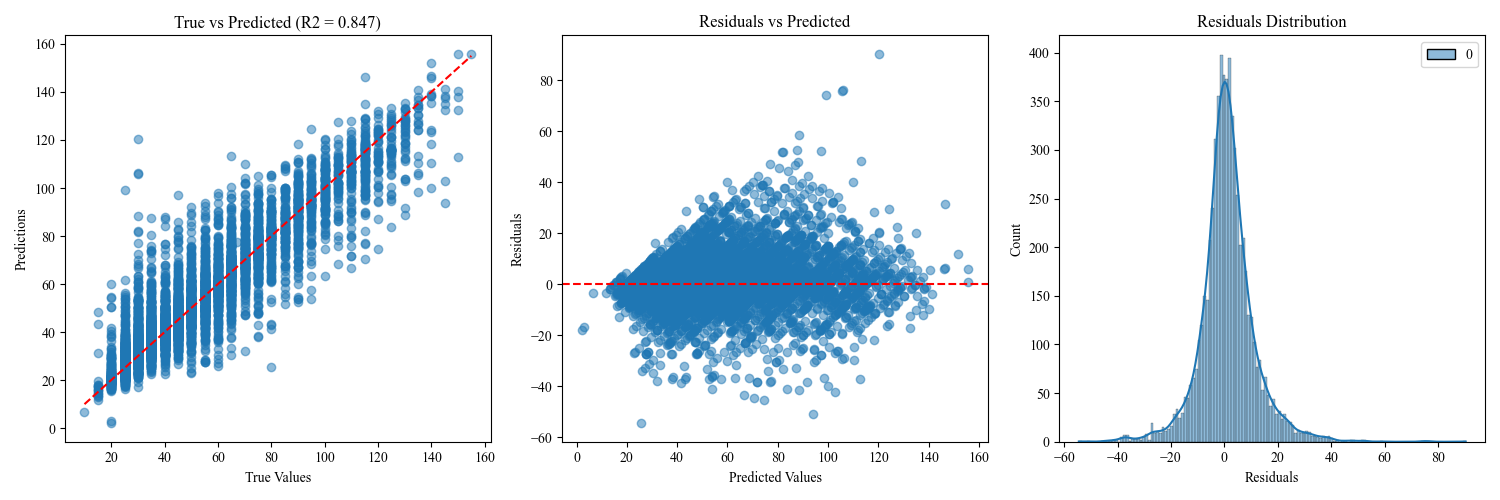} % Replace with your actual image file
    \caption{PIR-NODE Model Prediction Analysis Plots. }
    \label{fig:prediction_analysis}
\end{figure}

Figure \ref{fig:prediction_analysis} shows the prediction analysis plots of the PIR-NODE model, including a true vs. predicted value scatter plot, a residual vs. predicted value scatter plot, and a residual distribution histogram. The true vs. predicted value scatter plot shows that the predicted values are highly concentrated around the diagonal line, and the R² value is as high as 0.847, further validating the strong predictive power of the PIR-NODE model. The residual vs. predicted value scatter plot shows that the residual distribution is relatively random, without obvious patterns, indicating that the model does not have systematic bias. The residual distribution histogram corroborates the error distribution histogram in Figure \ref{fig:error_distribution}, again validating the statistical characteristics of the prediction errors.

\subsection{Ablation Study}
We conducted an ablation study to evaluate the contribution of each component of our PIR-NODE model. Table 2 presents the results, showing that both the residual connections and the physics-informed loss function contribute significantly to model performance.

\begin{table}[h]
\centering
\caption{Ablation study results}
\begin{tabular}{lccc}
\hline
Model Variant & RMSE (kt) & MAE (kt) & $R^2$ \\
\hline
Full PIR-NODE & 10.40 & 7.15 & 0.847 \\
Without Residual Connections & 12.18 & 8.43 & 0.802 \\
Without Physics-Informed Loss & 11.56 & 7.89 & 0.821 \\
Without Both & 13.72 & 9.95 & 0.732 \\
\hline
\end{tabular}
\end{table}

\section{Discussion}

The Physics-Informed Residual Neural Ordinary Differential Equation model (PIR-NODE) proposed in this study has achieved significant performance improvements in tropical cyclone intensity prediction. Experimental results demonstrate that the PIR-NODE model achieves an order-of-magnitude improvement in 24-hour intensity prediction accuracy compared to traditional statistical models and benchmark deep learning methods. The success of the PIR-NODE model is attributed to the following key factors:

\textbf{Advantages of Residual Neural ODEs:} Neural ODEs naturally model the continuous temporal evolution of TC intensity, while residual connections effectively enhance model depth and training stability, and preserve initial intensity information, ensuring the physical plausibility of prediction results.

\textbf{Physics-Informed Design Philosophy:} Through feature engineering and model architecture design, we explicitly incorporate physical knowledge into the model, guiding the model to learn dynamic patterns that are more consistent with physical laws, improving model prediction accuracy and generalization ability.

\textbf{Comprehensive Data Preprocessing and Rigorous Evaluation System:} High-quality data preprocessing ensures the data foundation for model training, and a rigorous evaluation system comprehensively validates the model's performance advantages.

Despite the significant progress achieved by the PIR-NODE model, this study still has certain limitations. For example, the current model primarily relies on features from the SHIPS dataset. Future work could consider integrating more modalities of data, such as satellite imagery, radar data, and ocean observation data, to more comprehensively characterize the physical state of TCs and environmental influences. Furthermore, the architecture and hyperparameters of the PIR-NODE model still have room for further optimization. For instance, more advanced residual block structures, adaptive integration methods, and more effective physics information integration strategies could be explored.

\section{Conclusion and Outlook}

This paper innovatively proposes a Physics-Informed Residual Neural Ordinary Differential Equation model (PIR-NODE) for tropical cyclone intensity prediction. Experimental results fully demonstrate the superiority of the PIR-NODE model in TC intensity prediction, achieving significant improvements in accuracy and robustness compared to traditional methods. The results of this study not only validate the immense potential of Neural ODEs in predicting complex geophysical systems but, more importantly, provide a new direction for developing more refined, physically interpretable, and higher-prediction-accuracy TC forecasting models in the future.

Future research will primarily focus on the following aspects:

\textbf{Multi-modal Data Fusion:} Explore how to effectively fuse multi-source heterogeneous data from satellites, radar, and oceans to construct a more comprehensive TC state representation and further improve model prediction accuracy.

\textbf{Physical Constraint Enhancement:} Investigate embedding deeper physical constraints (e.g., energy conservation, mass conservation) directly into the PIR-NODE model architecture to enhance model physical interpretability and prediction reliability \cite{Beucler2021, Kidger2020}.

\textbf{Uncertainty Quantification Prediction:} Develop PIR-NODE models capable of quantifying prediction uncertainty to provide forecasters with more comprehensive decision support information and enhance the risk assessment capability of TC forecasting.

\textbf{Model Lightweighting and Real-time Forecasting Applications:} Explore model compression and acceleration techniques to apply the PIR-NODE model to real-time TC intensity forecasting operations, serving disaster prevention and mitigation practices.

\section{Acknowledgements}

The authors gratefully acknowledge the supports by the Natural Science Foundation of Jiangsu Province(Grants No. BK20240700).

% 1

% 2

% 3

\end{document}